\documentclass[pdflatex,sn-mathphys,iicol]{sn-jnl}% Math and Physical Sciences Reference Style
\usepackage{amsmath}
\newcommand{\bra}[1]{\langle#1\rvert} % Bra
\newcommand{\ket}[1]{\lvert#1\rangle} % Ket
\jyear{2021}%

\theoremstyle{thmstyleone}%
\theoremstyle{thmstyletwo}%

\theoremstyle{thmstylethree}%

\begin{document}

\title[-]{Permutation Invariant Encodings for Quantum Machine Learning with Point Cloud Data}


\author*[1]{\fnm{Jamie} \sur{Heredge}}\email{heredgej@student.unimelb.edu.au}

\author[1,2]{\fnm{Charles} \sur{Hill}}\email{cdhill@unimelb.edu.au}

\author[1]{\fnm{Lloyd} \sur{Hollenberg}}\email{lloydch@unimelb.edu.au}

\author[1]{\fnm{Martin} \sur{Sevior}}\email{martines@unimelb.edu.au}

\affil*[1]{\orgdiv{School of Physics}, \orgname{University of Melbourne}, \orgaddress{\city{Parkville}, \postcode{VIC 3010}, \country{Australia}}}

\affil[2]{\orgdiv{School of Mathematics and Statistic}, \orgname{University of Melbourne}, \orgaddress{\city{Parkville}, \postcode{VIC 3010}, \country{Australia}}}


% A crucial step within any QML algorithm is the process of encoding classical data into a quantum circuit. So far the most promising implementations of QML have often involved quantum encodings designed using prior knowledge of the underlying data \cite{liu_rigorous_2021}. 

\abstract{
Quantum Computing offers a potentially powerful new method for performing Machine Learning. However, several Quantum Machine Learning techniques have been shown to exhibit poor generalisation as the number of qubits increases \cite{huang_power_2021}. We address this issue by demonstrating a permutation invariant quantum encoding method, which exhibits superior generalisation performance, and apply it to point cloud data (three-dimensional images composed of points). Point clouds naturally contain permutation symmetry with respect to the ordering of their points, making them a natural candidate for this technique. Our method captures this symmetry in a quantum encoding that contains an equal quantum superposition of all permutations and is therefore invariant under point order permutation. We test this encoding method in numerical simulations using a Quantum Support Vector Machine to classify point clouds drawn from either spherical or toroidal geometries. We show that a permutation invariant encoding improves in accuracy as the number of points contained in the point cloud increases, while non-invariant quantum encodings decrease in accuracy. This demonstrates that by implementing permutation invariance into the encoding, the model exhibits improved generalisation.
}

%%================================%%
%% Sample for structured abstract %%
%%================================%%

% \abstract{\textbf{Purpose:} The abstract serves both as a general introduction to the topic and as a brief, non-technical summary of the main results and their implications. The abstract must not include subheadings (unless expressly permitted in the journal's Instructions to Authors), equations or citations. As a guide the abstract should not exceed 200 words. Most journals do not set a hard limit however authors are advised to check the author instructions for the journal they are submitting to.
% 




\maketitle

\section{Introduction}

Quantum Machine Learning (QML) is a promising candidate for real-world applications of quantum technology \cite{biamonte_quantum_2017}. In recent years, a multitude of different techniques have been developed with the aim of using quantum computers to perform machine learning tasks \cite{zeguendry_quantum_2023, sajjan_quantum_2022} and QML techniques have been applied to a wide range of fields, including particle physics \cite{Heredge21, tuysuz_hybrid_2021}, medical data \cite{pregnolato_sars-cov-2_2023, azevedo_quantum_2022}, aerodynamics \cite{Yuan:2022jcw} and natural language processing \cite{meichanetzidis_grammar-aware_2023}. In many QML techniques, classical data is encoded into an exponentially larger quantum space where there is the possibility that the data may be separated more easily \cite{havlicek_supervised_2019}. This is a proposed source of quantum advantage over classical routines in the case that the quantum circuit performing the encoding cannot be efficiently simulated classically \cite{liu_rigorous_2021}. When trying to find suitable QML techniques for real-world data, it is important to use an advantageous encoding method for that data. A current active area of research is the search for methods to encode or represent different types of data in quantum devices, for example, finding techniques to represent two-dimensional images \cite{lisnichenko_quantum_2022, Anand2022, west2022}.

Many QML techniques, such as the Quantum Support Vector Machine (QSVM) \cite{havlicek_supervised_2019}, are motivated by the idea that encoding classical data into a higher dimensional quantum Hilbert space can simplify data classification. The circuit architecture used to encode classical data into a quantum state influences the class of functions that a QML algorithm can learn \cite{schuld_effect_2020}. Therefore, it is critical to find an optimal quantum encoding for the data in any QML method. However, some QML techniques do not generalise well as the number of qubits, and hence, the dimensionality of the Hilbert space increases \cite{huang_power_2021}. To improve generalisation, attempts have been made to reduce the expressivity of QML methods by introducing some form of inductive bias into the method. These attempts include projected kernels, where only a limited number of qubits are measured, thus projecting to a lower-dimensional space \cite{inductive_bias}, introducing inductive bias through the tuning of quantum kernel hyperparameters \cite{PhysRevA.106.042407}, and variational state-based approaches that are capable of encoding inductive biases directly into quantum states, which have been shown to improve generalisation in the context of learning zero-sum games \cite{bowles_contextuality_2023}. In this work, we present a method of introducing an inductive bias into a quantum encoding when the underlying data exhibits a permutation symmetry. By projecting our quantum-encoded state onto a symmetric subspace, this method exponentially reduces the encoding's dimensionality, leading to improved generalisation in our experiments.

\begin{figure}[h]%
\centering
  \includegraphics[width=1.0\linewidth]{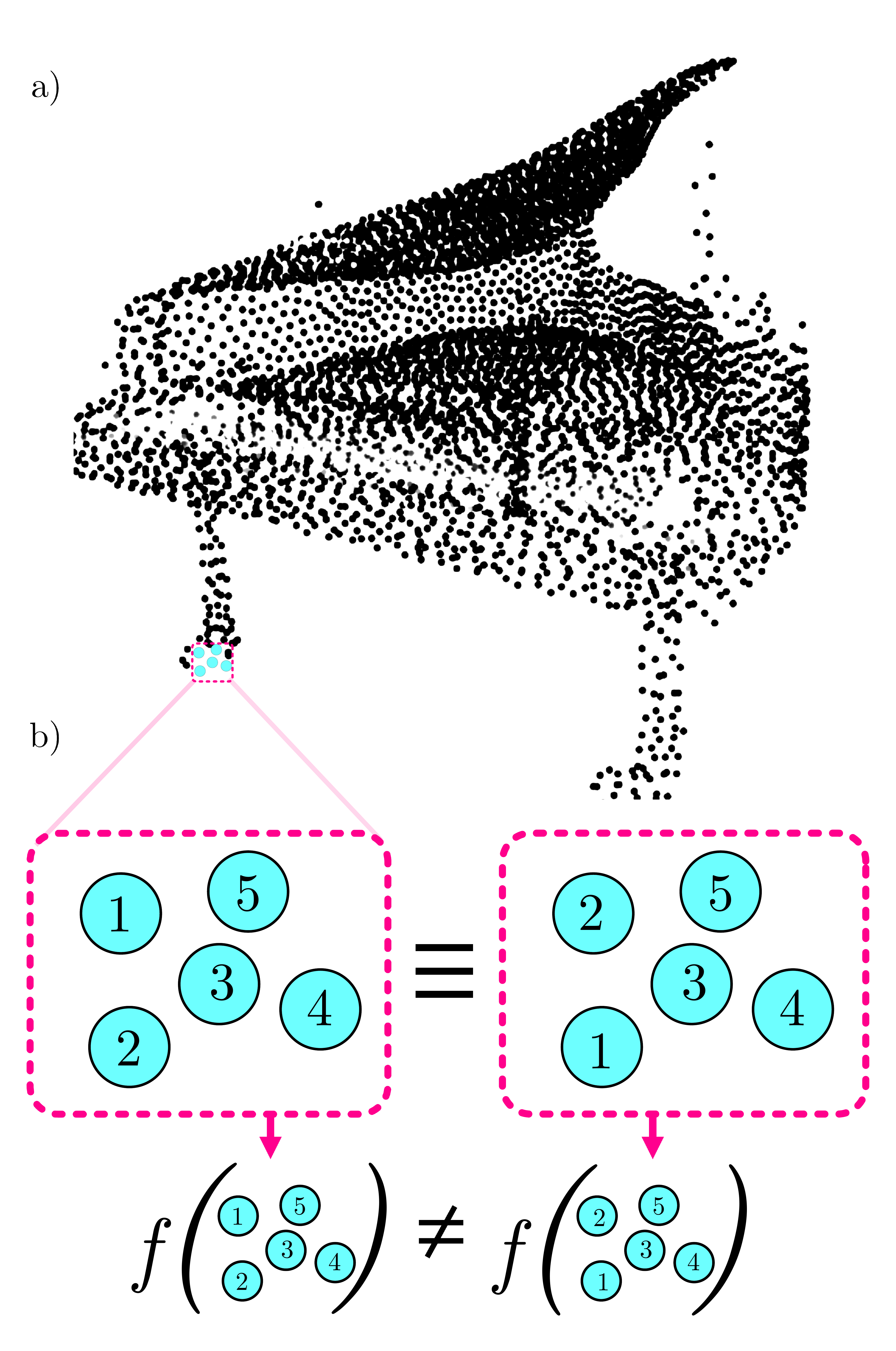}
\caption{a) Example point cloud generated using the Point-E demo by OpenAI using the prompt ``Grand Piano'' \cite{openAI_pointE}. Distinguishing between different objects could be a possible classification task that uses point cloud data. b) Demonstration of point permutation symmetry in the input for a point cloud. Changing the order of points in a point cloud does not have an effect on the point cloud itself. However, when stored as a classical input array in computer memory, exchanging point order produces a different array. Unless it has been purposely constructed otherwise, as is the case for PointNet \cite{pointnet}, a general machine learning classifier function, denoted by $f$, may produce a different classification output given a different order permutation of the point order in its input $f(x_1, y_1, z_1, x_2, y_2, z_2) \neq f(x_2, y_2, z_2, x_1, y_1, z_1)$.}
  \label{palm_tree_example}
\end{figure}

In this study, we consider point cloud data types, which are three-dimensional images that consist of a set of three-dimensional points. The point cloud may represent various objects (e.g. piano, car, tree) that need to be classified. This could be in the context of identifying pedestrians in self-driving vehicles \cite{Chen2021} or classifying different particle decay events in a high-energy physics experiment \cite{Mikuni_2021}. Point clouds are a natural data type to study when investigating the effects of permutation invariant machine learning methods. While they may also occasionally exhibit internal data-specific symmetries such as rotation or translation symmetry, we focus here on their point order permutation symmetry when they are input into a model and present a method of encoding this symmetry into a quantum state. Permutation symmetry is a property point cloud data possesses that is not normally captured in a classical input vector. There is no inherent ordering to the points in a point cloud. Therefore, if an order is assigned to the points in a point cloud, then it should be invariant under any permutation of these point labels. Classical computers are generally forced to assign an order to the points $\textbf{p}_i$ when they are input, since the data must be stored in an array in memory that has an intrinsic ordering to the points. Consider the input array $[\textbf{p}_1, \textbf{p}_2]$; exchanging two points in this array will produce a different array $[\textbf{p}_2, \textbf{p}_1]$ which may give a different result when input into a given algorithm. In general, a machine learning model classifier model $f([\textbf{p}_1, \textbf{p}_2])$, without purposeful construction, will return a different answer if given a different permutation of the same points in the input, $f([\textbf{p}_1, \textbf{p}_2]) \neq f([\textbf{p}_2, \textbf{p}_1])$, while in reality the point cloud would be physically unchanged by this reordering. By creating an encoding that is invariant to the permutation of point ordering, we can exponentially reduce the effective dimensionality of the encoded states in a manner that respects an underlying symmetry of the data. An example point cloud is shown in Figure~\ref{palm_tree_example} demonstrating the permutation invariance of points in the input.

There are other Machine Learning methods that aim to respect this permutation symmetry by building symmetric functions into the model, such as the max pool function in the classical PointNet \cite{pointnet} and the proposed quantum extension of PointNet \cite{Shi2020TrainingAQ}. Similarly, techniques in Geometric Quantum Machine Learning have shown it is possible to construct a variational circuit that respects qubit permutation (and hence point permutation if one were to encode a single point per qubit) \cite{meyer22, Nguyen22, schatzki_theoretical_2022, Kazi23}. The technique we discuss in this work differs in that we focus on implementing these symmetries into the encoding circuit, meaning that the classification part of the algorithm is free to take any form. Recent work suggests that permutation invariant operators, such as permutation equivariant variational circuits, may be classically tractable to simulate under certain conditions \cite{anschuetz_efficient_2023}. In our case, the quantum model does not necessarily need to be a permutation invariant operator, as the permutation invariance is captured in the encoding step.

\section{Permutation Invariant Encoding}

In this section, we theoretically outline the structure and properties of a permutation invariant quantum state within the context of point cloud data. Point clouds were chosen as a natural use case of this encoding; however, this could be applicable to any data that exhibits permutation symmetry. We consider a point cloud data input, denoted as $X$, to be an array of values in the form $X = [x_1, y_1, z_1, x_2, y_2, z_2,...,x_n, y_n, z_n]$. Each point cloud therefore consists of $n$ points, where a single point in the point cloud can be denoted as $\textbf{p}_i = [x_i, y_i, z_i]$. Each point is first encoded into a quantum state $\ket{\textbf{p}_i}$ using a quantum circuit $U$, consisting of $k$ qubits, that maps three-dimensional classical data $\textbf{p}_i$ to a $2^k$ dimensional quantum state, $U : \mathbf{R}^3 \xrightarrow{} \mathbf{R}^{2^k}$. We implement this gate on an initial $\ket{0}^{\otimes k}$ state such that $\ket{\textbf{p}_i} = U(\textbf{p}_i) \ket{0}^{\otimes k}$. While in our experiments $U$ was implemented using angle encoding, this general technique could be used for any encoding strategy. To enforce the point-exchange invariance, we construct a state that is in a symmetric superposition of all $\ket{\textbf{p}_i}$ permutation states. For a point cloud $X$ with only two points, $n = 2$, this can be represented as 
\begin{equation}\label{symm_state}
    \ket{X_s} = \mathcal{N}(\ket{\textbf{p}_1}\ket{\textbf{p}_2} + \ket{\textbf{p}_2}\ket{\textbf{p}_1}),
\end{equation} 
which is invariant under permutation of the order of the two points. Regarding the normalisation constant $\mathcal{N}$, it should be noted that depending on the data and the encoding method $U$, that the initial states $\ket{\textbf{p}_1}$ and $\ket{\textbf{p}_2}$ may or may not be orthogonal. Hence, the normalisation constant $\mathcal{N}$ for the 2 qubit case, in general, is
\begin{equation}
    \mathcal{N} = \frac{1}{\sqrt{2(1 +  \rvert \langle \textbf{p}_1 \rvert \textbf{p}_2 \rangle \rvert^2)}}.
\end{equation}
The point order invariant encoded state $\ket{X_s}$ can then be evaluated using techniques such as QSVM or passed to a variational method. As the input quantum state is now in a permutation invariant state, the quantum classification algorithm $g(\ket{X_s})$ is free to have any design and we are guaranteed to have permutation invariance under point order permutation as $g\Big( \mathcal{N}(\ket{\textbf{p}_1}\ket{\textbf{p}_2} + \ket{\textbf{p}_2}\ket{\textbf{p}_1}) \Big) = g\Big( \mathcal{N}(\ket{\textbf{p}_2}\ket{\textbf{p}_1} + \ket{\textbf{p}_1}\ket{\textbf{p}_2}) \Big)$ regardless of the structure of the quantum classification function $g$. This contrasts to a general machine learning classifier $f([\textbf{p}_1, \textbf{p}_2])$, accepting its input as a classical array, that may give different results depending on the order of points in the input such that $f([\textbf{p}_1, \textbf{p}_2]) \neq f([\textbf{p}_2, \textbf{p}_1])$. In the case of $n$ points, we wish to construct the symmetric state defined by
\begin{equation}
    \ket{X_s} = \mathcal{N}_n \sum_{\sigma \in S_n}\ket{\textbf{p}_{\sigma_1}}\ket{\textbf{p}_{\sigma_2}}...\ket{\textbf{p}_{\sigma_n}},
\end{equation}
where we sum over all permutations in the symmetric group $S_n$. This symmetric state will be identical under any permutation of points.

By combining the quantum states into a symmetric superposition state, we utilise the inherent quantum property of state superposition to implement an underlying symmetry of the data structure into the encoding. This symmetry exploitation allows for a reduction in the expressivity of the encoding, by exponentially reducing the effective dimensionality of the state. This can be demonstrated by considering a three qubit state. In general a three qubit state is $2^3 = 8$ dimensional and can be written as
\begin{align}
\begin{split}\label{three_qubit_dim_eqn}
    \ket{\psi} & = \alpha_0 \ket{0}\ket{0}\ket{0} \\
    &+\alpha_1 \ket{1}\ket{0}\ket{0} + \alpha_2 \ket{0}\ket{1}\ket{0} + \alpha_3 \ket{0}\ket{0}\ket{1} \\ 
    &+\alpha_4 \ket{1}\ket{1}\ket{0} + \alpha_5 \ket{1}\ket{0}\ket{1} + \alpha_6 \ket{0}\ket{1}\ket{1} \\ 
    &+\alpha_7 \ket{1}\ket{1}\ket{1}.
\end{split}
\end{align}
If we now insist that the state $\ket{\psi}$ is fully symmetric with respect to its qubits, then by exchanging qubits in Equation~\ref{three_qubit_dim_eqn} and ensuring that the state remains unchanged under this action, it can be seen that $\alpha_1 = \alpha_2 = \alpha_3$ and $\alpha_4 = \alpha_5 = \alpha_6$ \cite{barenco1996}. Hence, a three-qubit symmetric quantum state is effectively $4$ dimensional with the following basis states
\begin{align}
\begin{split}\label{three_qubit_dim_eqn_sym}
    \ket{\psi} & = \beta_0 \ket{0}\ket{0}\ket{0} \\
    &+\beta_1 (\ket{1}\ket{0}\ket{0} + \ket{0}\ket{1}\ket{0} + \ket{0}\ket{0}\ket{1}) \\ 
    &+\beta_2 (\ket{1}\ket{1}\ket{0} + \ket{1}\ket{0}\ket{1} + \ket{0}\ket{1}\ket{1}) \\ 
    &+\beta_3 \ket{1}\ket{1}\ket{1}.
\end{split}
\end{align}
It follows that an $n$ qubit system that is permutation invariant with respect to its qubits has dimension $n+1$, which is exponentially smaller than $2^n$ \cite{barenco1996}. In the general case, where each initial state $\textbf{p}_i$ has $k$ qubits it has been shown that the dimension of this symmetric state is
\begin{equation}
    ^{n+2^k-1}C_{2^k - 1} = \frac{1}{(2^k-1)!}n^{2^k - 1} + O(n^{2^k - 2}),
\end{equation}
which exhibits polynomial scaling in $n$, in contrast to the general case where the dimension is $(2^k)^n$ and the dimension scales exponentially \cite{barenco1996}. Note that if the data requires, such as in cases of underfitting the training data, we retain the ability to reintroduce some expressivity through breaking of the symmetry or by increasing the number of qubits $k$ used per point.

For the theoretical results in this work we used Qiskit $\textit{statevector\_simulator}$ along with an analytical symmetrisation process as described in Algorithm~\ref{algo1} that mathematically constructs the permutation invariant quantum states. A discussion around possible implementations of this procedure on a real quantum machine is the focus of Section~\ref{sec_circuit}. 

\begin{algorithm}
\caption{Encode point cloud data with point permutation invariance using Qiskit $\textit{statevector\_simulator}$ and analytical symmetrisation. This algorithm demonstrates the mathematical structure of the encoding intuitively at the cost of being computationally inefficient by containing $\mathcal{O}(n!)$ classical computations.}\label{algo1}
\begin{algorithmic}[1]
\Require{Array $X$ of $n$ points $\textbf{p}_i$ where $0 < i \leq n$}
\For {$i$ in $(0,n]$}
\State $U(\textbf{p}_i)\ket{0}^{\otimes k} = \ket{\textbf{p}_i}$ on separate registers
\State Evaluate $\ket{\textbf{p}_i}$ using $\textit{statevector\_simulator}$
\EndFor
\State Initialise empty symmetric statevector $\ket{X_s}$
\For {all permutations $\sigma$ in symmetric group $ S_n$}
\State $\ket{X_{\sigma}} \Leftarrow \ket{\textbf{p}_{\sigma_1}}$
\State $j \Leftarrow 2$
\While {$j \leq n$}
\State $\ket{X_{\sigma}} \Leftarrow \ket{X_{\sigma}} \otimes \ket{\textbf{p}_{\sigma_j}}$
\State $j \Leftarrow j + 1$
\EndWhile
\State $\ket{X_s} \Leftarrow \ket{X_s} + \ket{X_{\sigma}}$
\EndFor
\State Normalise $\ket{X_s} \Leftarrow \frac{\ket{X_s}}{\sqrt{\langle X_s \rvert X_s \rangle}}$
\State Return $\ket{X_s}$
\Ensure Permutation invariant statevector $\ket{X_s}$
\end{algorithmic}
\end{algorithm}

\section{Sphere and Torus Classification}\label{section_sphere}

We compare the performance of various quantum and classical machine learning techniques when applied to the classification of two point cloud data distributions: a sphere and a torus. To generate the data, we randomly sample $n$ points from the surface of each shape to form a point cloud for each distribution. This sampling process is repeated until we have $N$ point cloud samples in total, dividing the resulting data into training and testing sets with 80\% and 20\% of the data, respectively. The performance of various models is then evaluated on the testing set. The entire process is then repeated with new randomly generated datasets and we record the average test accuracy for 10 repeated experiments.

\begin{figure}[h]%
\centering
\includegraphics[width=0.95\linewidth]{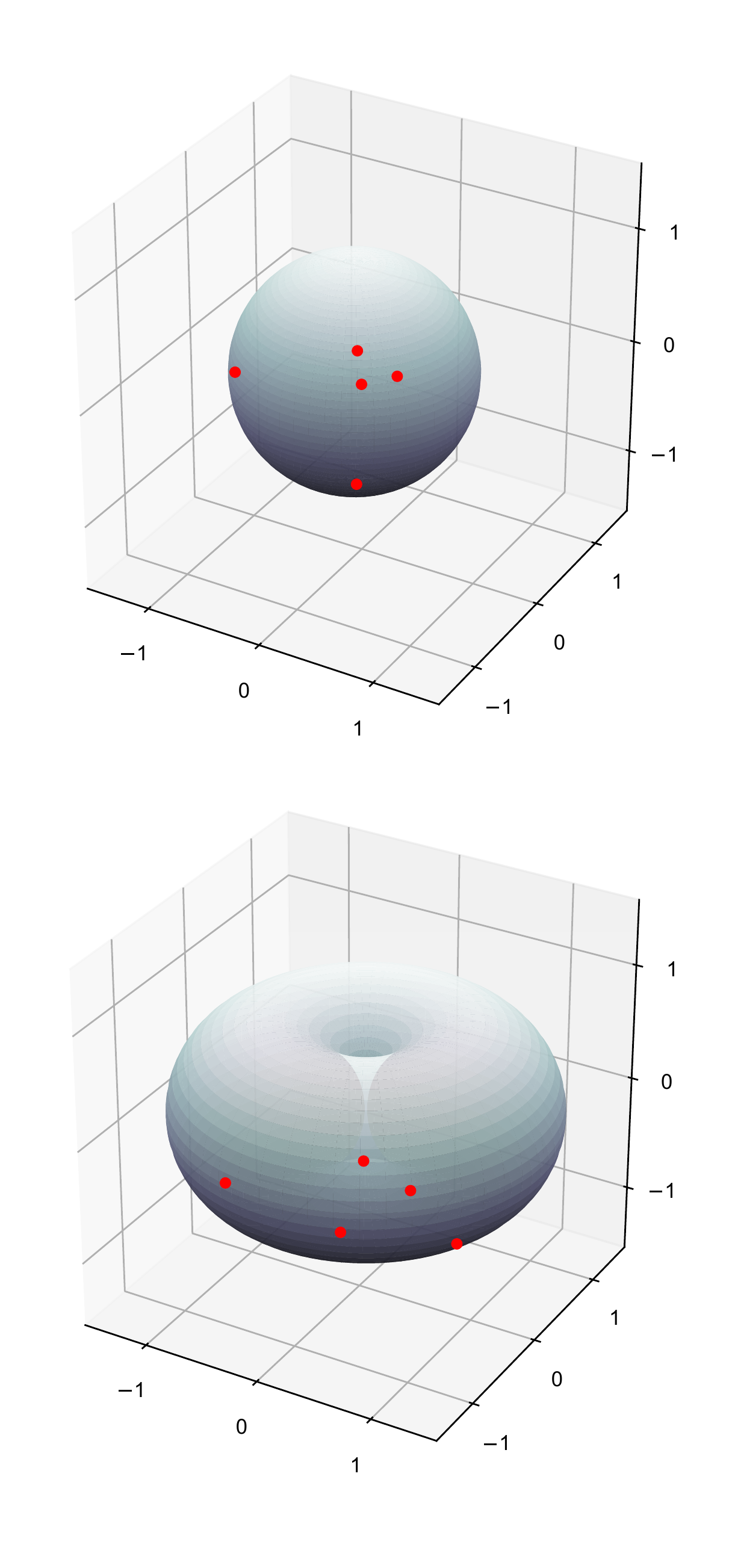}
\caption{(left) Sphere distribution and (right) torus distribution surfaces from which points are sampled to form point clouds. The red points in each figure represent an example point cloud with $n = 5$ points drawn from the corresponding distribution. Each dataset is generated by randomly sampling these distributions to create a set of $N$ different point clouds.}\label{dist_example}
\end{figure}

The sphere and torus distributions were both centred at the origin. To ensure that the sphere and torus distributions are as similar as possible, the torus was scaled such that the average magnitude of the points that lie on the torus distribution surface matches the radius of the sphere. An illustration of the two distributions and an example point cloud sample is shown in Figure~\ref{dist_example}. All data was normalised to be in the range $\frac{\pi}{2}$ to $-\frac{\pi}{2}$ to allow it to be encoded as rotation angles.

\begin{figure}[h]%
\centering
  \includegraphics[width=0.9\linewidth]{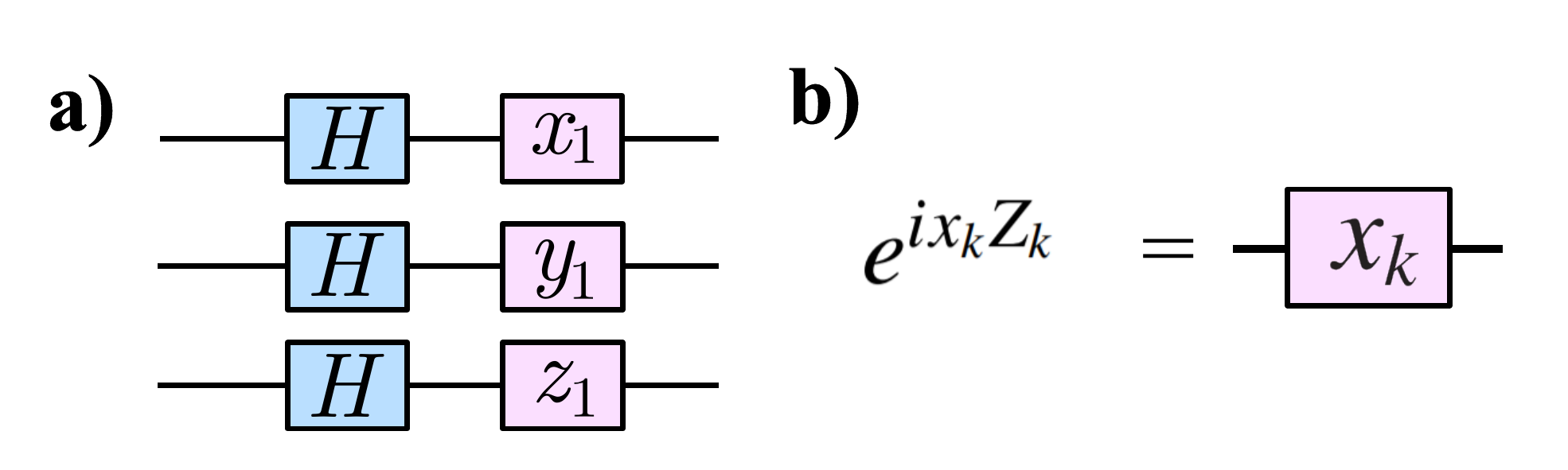}
  \caption{a) Single layer of the best performing point encoding circuit $U_{\alpha}$ found for the sphere / torus dataset during our investigation, which is used for results titled Permutation Invariant QSVM (Best). b) Pink boxes represent parameterised Z rotation gates.}
  \label{best_encoding}
\end{figure}

\begin{figure}[h]%
\centering
  \includegraphics[width=0.9\linewidth]{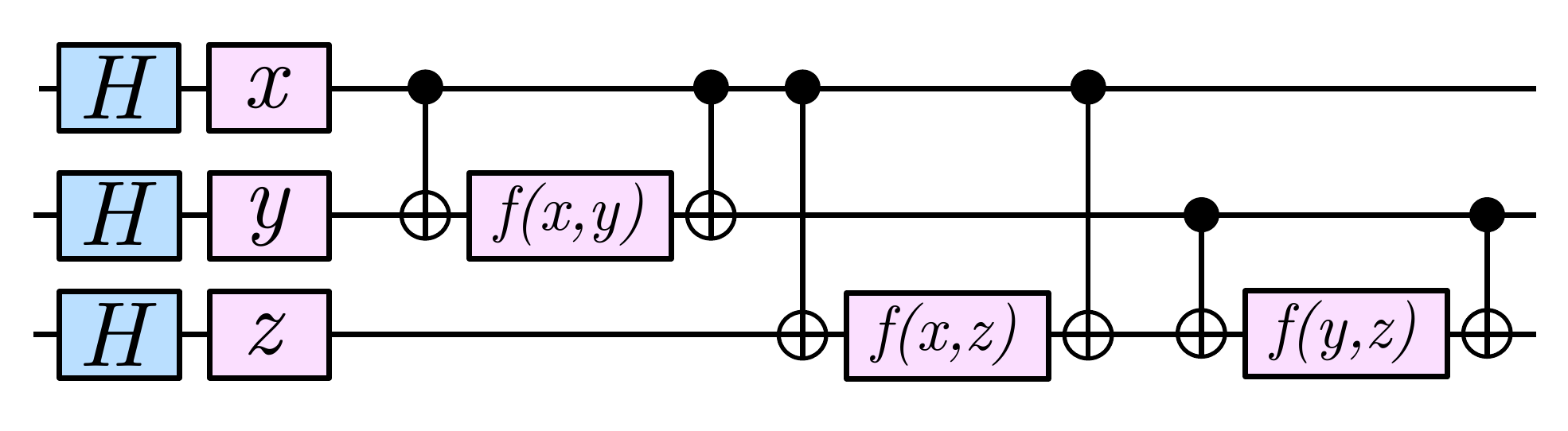}
  \caption{a) The point encoding circuit $U_{\beta}$, that uses the IQP encoding \cite{havlicek_supervised_2019}, which corresponds to results titled Permutation Invariant IQP QSVM. The entanglement function was defined as $f(x,y)=\frac{1}{\pi}(\pi-x)(\pi-y)$.}
  \label{iqp_encoding}
\end{figure}

We tested a variety of quantum and classical techniques, summarised in Table~\ref{comparisons}, where some of the algorithms contain permutation invariance and others do not. In order to focus entirely on the encoding method, without having to consider the structure of a variational component, we used a QSVM to classify the data with various choices of encoding circuits. For the order permutation invariant encoding we tested several different point encoding circuits $U$ for encoding the individual points, showing results for the best-performing circuit denoted $U_{\alpha}$ alongside a more generic point encoding circuit that uses the Instantaneous Quantum Polynomial IQP encoding \cite{havlicek_supervised_2019}, denoted by $U_{\beta}$. We provide a brief description of the different methods reported:
\begin{itemize}
  \item Permutation Invariant QSVM (Best) - A QSVM with the permutation invariant encoding method using the point encoding circuit $U_{\alpha}$ that provided the highest accuracy, as shown in Figure~\ref{best_encoding}.
  \item Permutation Invariant IQP QSVM - A QSVM with the permutation invariant encoding method where the point encoding circuit $U_{\beta}$ is an IQP encoding, as shown in Figure~\ref{iqp_encoding}. This is to provide a more fair comparison between the regular IQP encoding and the invariant encoding.
  \item IQP Encoding QSVM - A QSVM using the IQP encoding applied to all variables in the input (without permutation invariance) as described by Havlicek et al. \cite{havlicek_supervised_2019}. 
  \item PointNet - Classical point cloud classifier algorithm utilising neural networks with a symmetric max pool function to ensure point order permutation invariance \cite{pointnet}. PointNet was run over 100 training epochs.
  \item RBF Kernel SVM - Classical SVM using the Radial Basis Function (RBF) kernel. Hyperparameters were optimised using grid search over a cross-validation set. 
\end{itemize}

 \begin{table*}[h]
 \caption{Summary of the various algorithms tested indicating whether they contain permutation invariance in their design and whether they are quantum or classical approaches.}\label{comparisons}%
\begin{center}
\begin{tabular}{|l|l|l|}
\hline
 & Quantum & Classical \\\hline
Permutation Invariant & Permutation Invariant QSVM (Best) & PointNet \\
 & Permutation Invariant IQP QSVM &  \\\hline
Non-invariant & IQP Encoding QSVM & RBF Kernel SVM \\\hline
\end{tabular}
\end{center}
\end{table*}

\begin{figure}[h]%
\centering
\includegraphics[width=1.1\linewidth]{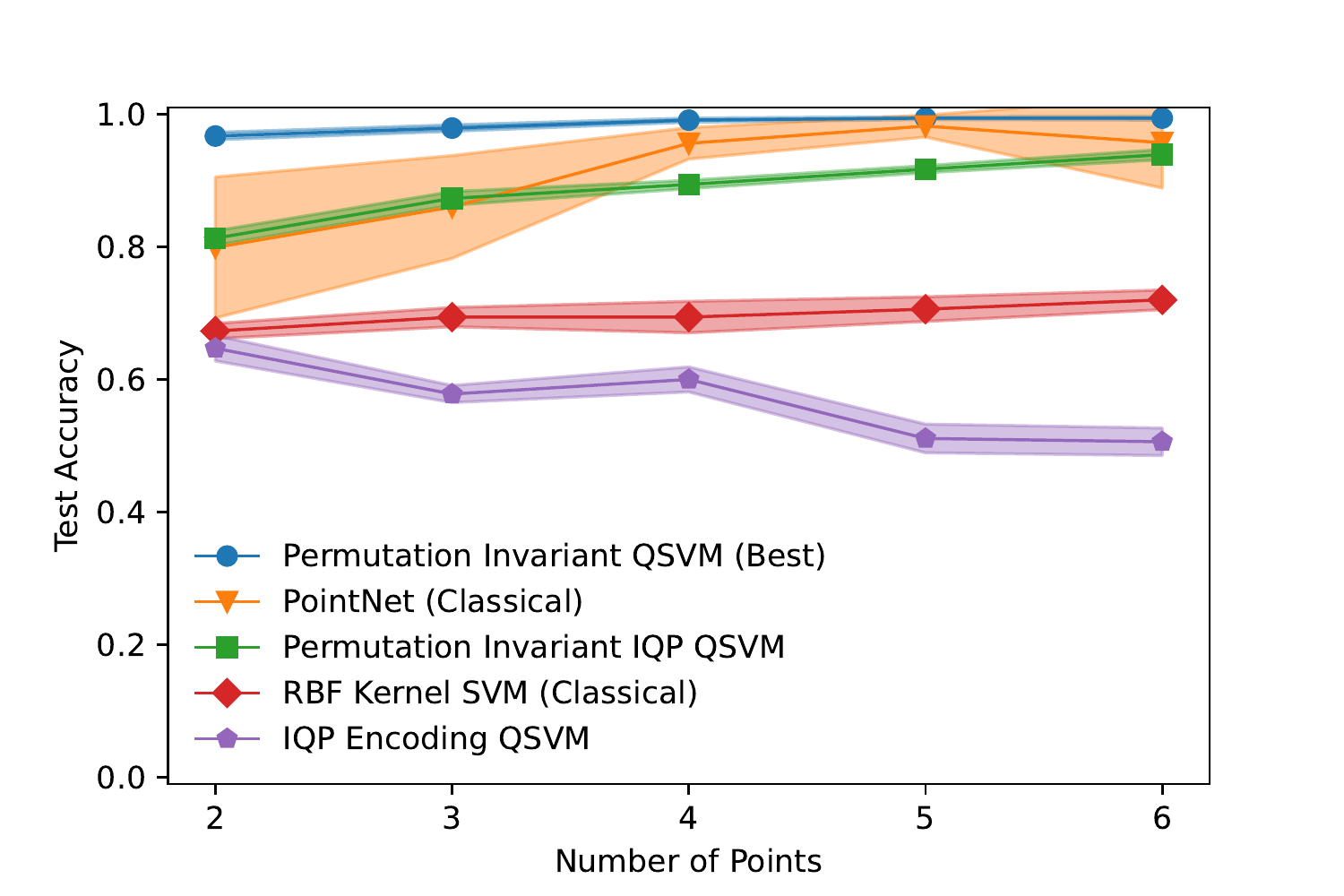}
\caption{The average accuracy over 10 repeated experiments as the number of points in each point cloud increases. Shaded regions indicate the error bounds on the average accuracy. Each experiment consists of a random dataset sample of 500 point clouds. Each point cloud is generated by randomly sampling a number of points from either the torus or the sphere distribution. The training and testing data contained 80\% and 20\% of the total data respectively.}\label{scaling_points}
\end{figure}

Our results were obtained from noiseless simulations using Qiskit \cite{qiskit}. Figure~\ref{scaling_points} displays how the various techniques scale as the number of points in the point cloud increases. Although more points provide more information, the non-invariant IQP QSVM's performance decreases as the number of points increases. This finding is consistent with previous research indicating that generic QSVM methods may struggle to generalise as the number of qubits increases \cite{huang_power_2021}. In contrast, the Permutation Invariant IQP Encoding exhibits an improvement in accuracy as the number of points increases. This result indicates that the symmetrisation technique may help prevent poor scaling due to the reduced expressivity in the encoding. Additionally, our tests reveal that using the best encoding circuit $U_{\alpha}$ produces a classifier that can outperform the classical PointNet algorithm for this dataset. This is further demonstrated in the results depicted in Table~\ref{results} which shows that for small datasets, the permutation-invariant quantum encoding outperforms both non-invariant quantum/classical methods and the permutation-invariant classical method PointNet.

 \begin{table}[h]
\begin{center}
\begin{minipage}{174pt}
\caption{The average accuracy over 10 repeated experiments. Each experiment consists of a random dataset sample of 100 point clouds consisting of 3 points each run using Qiskit \textit{statevector\_simulator}. The training and testing data contained 80\% and 20\% of the total data, respectively.}\label{results}%
\begin{tabular}{@{}ll@{}}
\toprule
 Algorithm & Accuracy \\
\midrule
 Permutation Invariant (Best) & 0.950 $\pm$ 0.015 \\
 PointNet (Classical) & 0.765 $\pm$ 0.038 \\
 %Particle-inspired QSVM  &	0.65 $\pm$ 0.028 \\ 
 RBF Kernel SVM (Classical) & 0.595 $\pm$ 0.022 \\ 
 Permutation Invariant IQP QSVM & 0.840 $\pm$ 0.038 \\
 IQP Encoding QSVM & 0.505 $\pm$ 0.036 \\
\botrule
\end{tabular}
\end{minipage}
\end{center}
\end{table}

\begin{figure}[h]%
\centering
\includegraphics[width=1.1\linewidth]{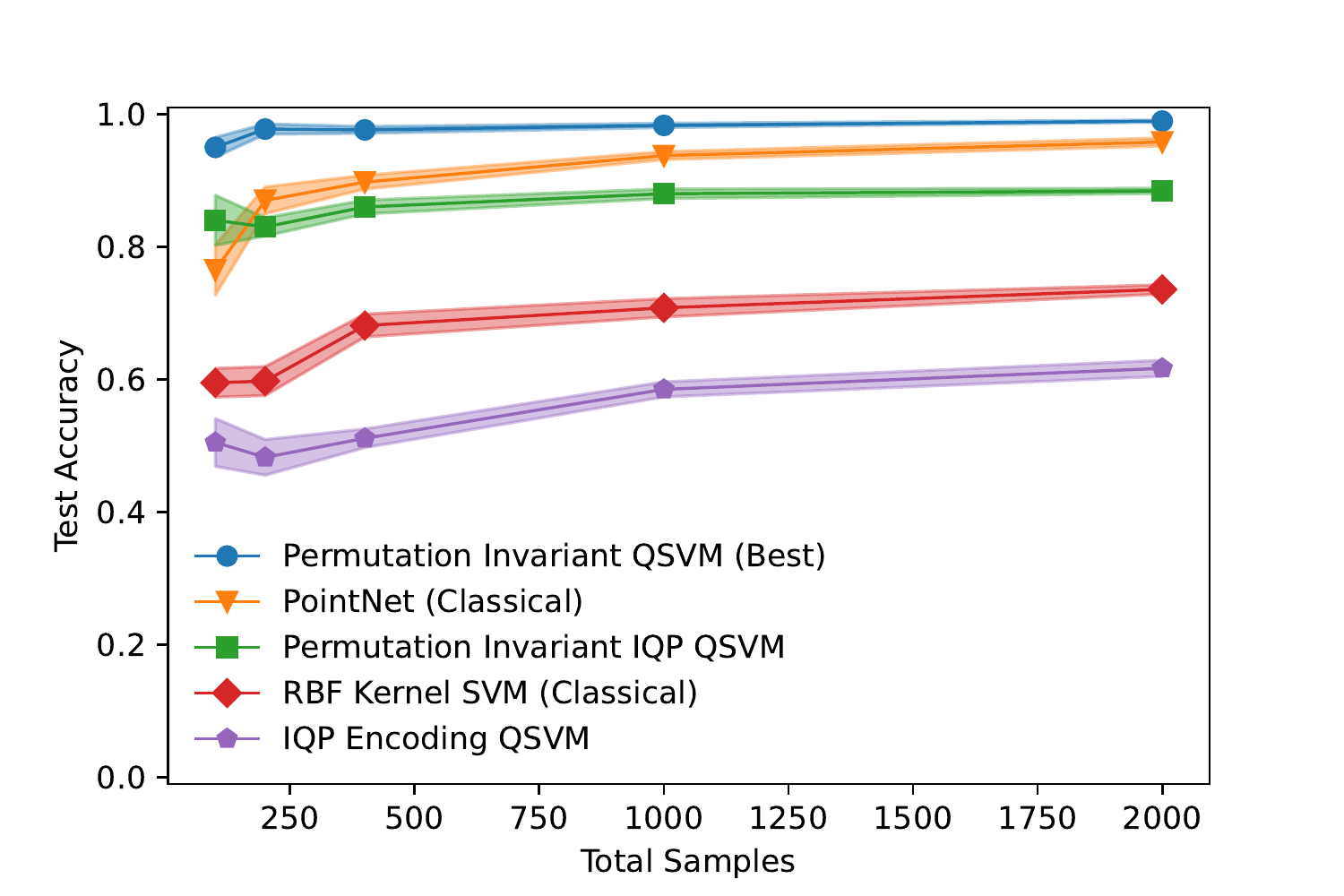}
\caption{The average accuracy over 10 repeated experiments as the number of samples in the training and testing dataset increases. Shaded regions indicate the error bounds on the average accuracy. Each experiment consists of a random dataset sample of point clouds. Each point cloud is generated by randomly sampling 3 points from either the torus or the sphere distribution. The training and testing data contained 80\% and 20\% of the total data respectively.}\label{scaling_sample}
\end{figure}

 The effect of increasing the size of the data sample is shown in Figure~\ref{scaling_sample}. In this case more data is available, but the number of points, and thus qubits is fixed. All algorithms tested generally improve with more data samples, as expected. Comparing again to Figure~\ref{scaling_points} shows that the IQP encoding, while improving with more data samples generally, is specifically struggling when there is an increase in qubits. This problem is not apparent with the permutation-invariant encodings.

 For larger data samples, PointNet starts to approach the accuracy of the permutation invariant QSVM encoding. It is worth noting that PointNet uses deep neural networks consisting of a total of 3.5 million parameters that can be better utilised with a large amount of training data. There are also additional aspects to the PointNet algorithm that tackle geometric symmetries, such as rotational invariance, in point cloud data that have not been implemented in this quantum approach. Implementing these geometric symmetries into the encoding could be a subject of further investigation. 

\section{Circuit Implementation}\label{sec_circuit}

\begin{figure}[h]%
\centering
\includegraphics[width=0.9\linewidth]{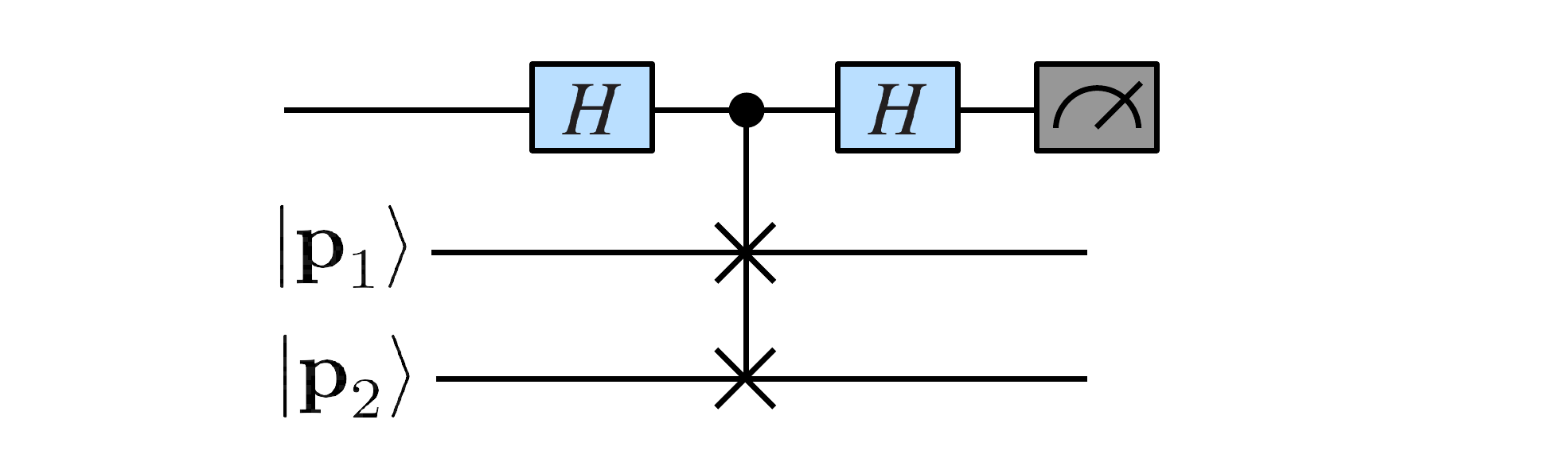}
\caption{Permutation symmetrisation circuit for two points. The circuit consists of a $\ket{p_1}$ state, a $\ket{p_2}$ state, and an ancilla qubit that performs a controlled swap operation. The final state of this circuit is given by $\ket{X} = \frac{1}{2}\ket{0}(\ket{\textbf{p}_1}\ket{\textbf{p}_2} + \ket{\textbf{p}_2}\ket{\textbf{p}_1}) + \frac{1}{2}\ket{1}(\ket{\textbf{p}_1}\ket{\textbf{p}_2} - \ket{\textbf{p}_2}\ket{\textbf{p}_1})$. This state is symmetric when the ancilla qubit is measured in the state $\ket{0}$ and anti-symmetric when it is measured in the state $\ket{1}$. By measuring the ancilla qubit and discarding any measurements in the state $\ket{1}$, we are left with the desired symmetric quantum state $\ket{X_s} = \mathcal{N}(\ket{\textbf{p}_1}\ket{\textbf{p}_2} + \ket{\textbf{p}_2}\ket{\textbf{p}_1})$.}\label{fig_2_qubit_example}
\end{figure}

The results in Section~\ref{section_sphere} suggest that permutation invariant encodings could be useful for point cloud data. These results were however created using analytical simulations, which required $\mathcal{O}(n!)$ classical operations. We now present a discussion on the practical aspects of implementing them the encoding directly onto a real quantum device. It has been shown that the symmetrisation process $\ket{\textbf{p}_1}\ket{\textbf{p}_2} \Rightarrow \mathcal{N}(\ket{\textbf{p}_1}\ket{\textbf{p}_2} + \ket{\textbf{p}_2}\ket{\textbf{p}_1})$ cannot be done perfectly by a unitary transformation \cite{buzek_optimal_2000}. However, it can be implemented in a probabilistic manner using controlled swap gates and ancilla qubits \cite{barenco1996}. For a point cloud data sample that contains only two points, we start with each point having been encoded in a separate quantum state $\ket{\textbf{p}_1} = U(\textbf{p}_1) \ket{0}^{\otimes k}$ and $\ket{\textbf{p}_2} = U(\textbf{p}_2) \ket{0}^{\otimes k}$, using an encoding circuit $U$. The symmetrisation procedure needs to produce the state 
\begin{equation*}
    \ket{X_s} = \mathcal{N}(\ket{\textbf{p}_1}\ket{\textbf{p}_2} + \ket{\textbf{p}_2}\ket{\textbf{p}_1}),
\end{equation*}
 such that exchanging the ordering of the two points in the input will leave this new quantum state invariant. This can be achieved in the two-qubit case by preparing an ancilla qubit using a Hadamard gate in the state $\frac{1}{\sqrt{2}}(\ket{0} + \ket{1})$ and using it to apply a controlled swap gate to the two input states $\ket{\textbf{p}_1}$ and $\ket{\textbf{p}_2}$, followed by another Hadamard gate applied to the ancilla qubit. This action leaves the system in the state
\begin{equation}\label{2_qubit_final_state_prep}
    \frac{1}{2}\ket{0}(\ket{\textbf{p}_1}\ket{\textbf{p}_2} + \ket{\textbf{p}_2}\ket{\textbf{p}_1}) + \frac{1}{2}\ket{1}((\ket{\textbf{p}_1}\ket{\textbf{p}_2} - \ket{\textbf{p}_2}\ket{\textbf{p}_1}),
\end{equation}
which contains both the permutation symmetrised and anti-symmetrised states. By measuring the ancilla qubit and discarding any result when the ancilla qubit is in the $\ket{1}$ state (corresponding to the anti-symmetric state), we arrive at the permutation symmetrised state whenever the ancilla qubit is measured in the $\ket{0}$ state. Inspecting Equation~\ref{2_qubit_final_state_prep} it can be seen that the probability of measuring the ancilla qubit in the desired $\ket{0}$ state is $\frac{1}{2}(1 + \rvert \langle \textbf{p}_1 \rvert \textbf{p}_2 \rangle \rvert^2)$ \cite{buzek_optimal_2000}, which means that in the worst-case scenario, when the input states are orthogonal, the probability is $\frac{1}{2}$. This symmetrisation procedure for a two-qubit system is illustrated in Figure~\ref{fig_2_qubit_example}. 

\begin{figure}[h]%
\centering
\includegraphics[width=0.9\linewidth]{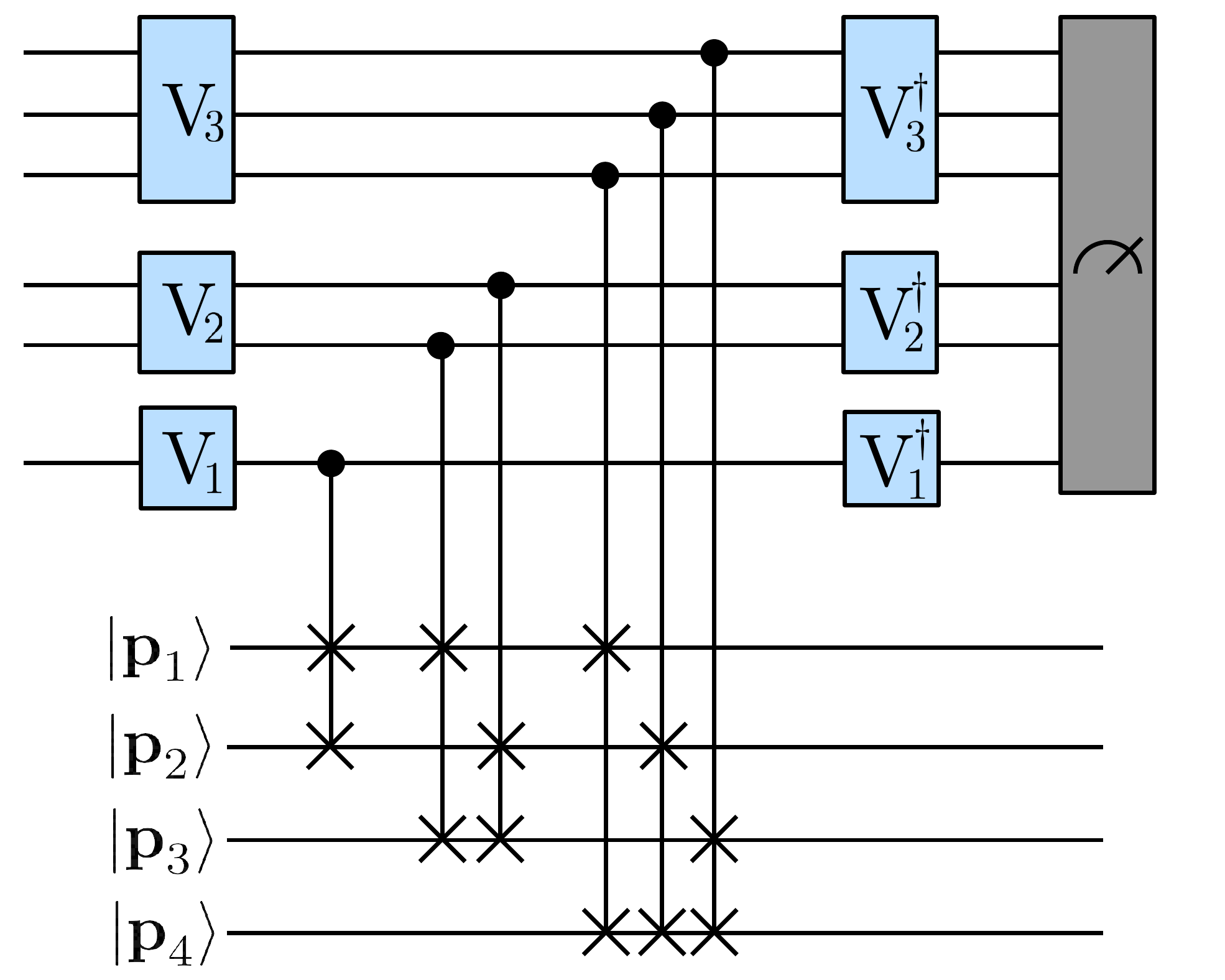}
\caption{Generalisation of the permutation symmetrisation process, as proposed by Barenco et al. \cite{barenco1996}, applied to a circuit containing four input states. The unitary operators $V_3$ perform a transformation on the ancilla qubits such that they are in the state $\frac{1}{\sqrt{4}}(\ket{000} + \ket{100} + \ket{010} + \ket{001})$. Similarly $V_2$ prepares $\frac{1}{\sqrt{3}}(\ket{00} + \ket{10} + \ket{01})$ and $V_1$ prepares $\frac{1}{\sqrt{2}}(\ket{0} + \ket{1})$. Through implementing controlled swap gates, these ancilla qubits will perform every possible permutation of the input states. This results in an equal superposition of every permutation of the input states after the controlled swap gates are applied and the ancilla qubits, after having $V^\dagger$ applied, are measured to be in the $\ket{000000}$ state. This process can be extended to any number of input states.}\label{fig_4_qubit_example}
\end{figure}

\begin{figure}[h]%
\centering
\includegraphics[width=0.9\linewidth]{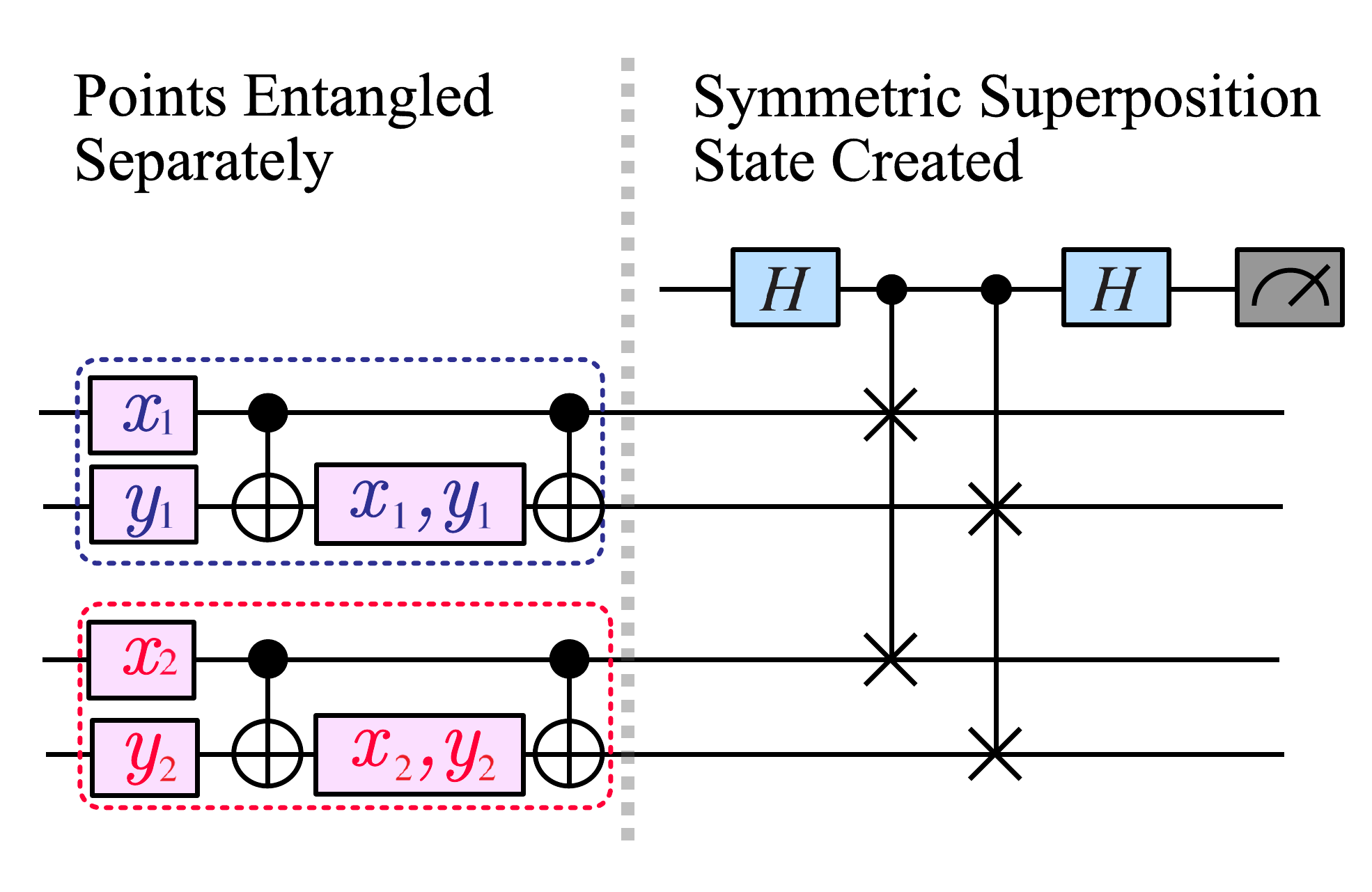}
\caption{Permutation symmetrisation of a 2-dimensional point cloud consisting of two points. On the left side, we first create an entangled state for each point using an encoding function $U$. On the right-hand side, we implement the symmetrisation process for a two-state system. If we discard any measurements in which the ancilla qubit is in the state $\ket{1}$, we will have prepared a symmetric encoding of this point cloud.}\label{point_cloud_sym_example}
\end{figure}

This procedure can be generalised to $n$ qubits using the technique outlined by Barenco et al. \cite{barenco1996}, which involves the iterative application of controlled swap symmetrisation operations using ancilla qubits. In this technique, the unitary operator $V_f$ prepares $f$ ancilla qubits into the state
\begin{equation}\label{plus_state_form}
    \frac{1}{\sqrt{f + 1}}(\ket{00...0} + \ket{10...0} + \ket{01...0} + ... + \ket{00...1}).
\end{equation}
This is carried out for $f=1$ to $f=n-1$ so that we have in total $c = \frac{1}{2}n(n-1)$ ancilla qubits. The ancilla qubits are then used to apply controlled swap gates onto the input states, followed by an application of the inverse of the unitary operations $V_f^\dagger$ on the ancilla qubits before their measurement. This recursively applies every possible qubit permutation to the input states, resulting in the desired symmetric superposition state when the ancilla qubits are measured in the state $\ket{0}^{\otimes c}$. An example circuit for a four-qubit scenario is shown in Figure~\ref{fig_4_qubit_example}. When each state is composed of multiple qubits, no extra ancilla qubits are necessary. Instead, additional controlled swap gates are applied to the extra qubits in the same manner \cite{buzek_optimal_2000}. This is demonstrated in Figure~\ref{point_cloud_sym_example}, which shows the symmetrisation of a two-dimensional point cloud consisting of only two points.

One drawback of this probabilistic implementation is the need to discard any states when the ancilla qubits are not in the state $\ket{0}^{\otimes c}$. The probability of this happening depends on the states themselves, with the probability being $1$ if the input states happen to be all identical, and decreasing as the overlap between states decreases. The probability of measuring the ancilla qubit in the state $\ket{0}$ for the case of two qubits can be shown to be $\frac{1}{2}(1 + \rvert \langle \textbf{p}_1 \rvert \textbf{p}_2 \rangle \rvert^2)$ \cite{buzek_optimal_2000}. In this work we utilised angle encoding for the point encoding circuit $U$ to produce the input states $\ket{\textbf{p}_1}$ and $\ket{\textbf{p}_2}$. This technique results in a relatively low probability for two specific data points to be orthogonal. As a result, the average probability of success remains relatively high even with an increase in the number of data points $n$. Additional details and supporting evidence for this claim are presented in Appendix~\ref{apd_success_prob}.

Although this method would be viable for small point clouds, the scaling of the probabilistic ancilla-based symmetrisation becomes problematic when a point cloud consists of a large number of points. This is because as the number of ancilla qubits increases, so does the number of gates in the circuit, as well as the number of states that need to be discarded. An alternative approach could be to approximately implement these states by producing states of the form $\ket{X_e} \approx \mathcal{N}_e(\ket{\textbf{p}_1}\ket{\textbf{p}_2} + \ket{\textbf{p}_2}\ket{\textbf{p}_1})$. This could potentially be achieved using techniques such as genetic state preparation algorithms \cite{creevey23} or Quantum Generative Adversarial Networks \cite{zoufal_quantum_2019}. Creating approximate symmetric states within some error $\epsilon$ could even be used to introduce a parameterised symmetry-breaking term that may help fine-tune the model, as has been shown to be useful in some variational quantum eigensolvers \cite{park21}. A parameterised method to increase the dimensionality of the problem through symmetry breaking would allow control over the expressibility of the encoding, as well as potentially making the encoding harder to simulate classically.

Alternatively, there is also the possibility of not discarding any states and instead using other superposition states (mixtures of anti-symmetric and symmetric permutation terms with respect to different points). For example, in the two-point case, one could accept the anti-symmetric state $\ket{X_a} = \mathcal{N}_a (\ket{\textbf{p}_1}\ket{\textbf{p}_2} - \ket{\textbf{p}_2}\ket{\textbf{p}_1})$. This state will exhibit a quasi-symmetry through the fact that permuting points results in a phase shift in the quantum state, but keeps their relative magnitudes intact. The effect that quasi-symmetric or approximately symmetric encodings would have on a QML technique, especially in situations where the model may be able to learn to overcome this peculiarity, is a possible subject for further research. 

\section{Conclusion}

This study presents a method for encoding point cloud data into a quantum state that is invariant under point order permutation, using a symmetrisation process to create a quantum superposition of all order permutations of the points. This exponentially reduces the dimensionality of the encoding, leading to an encoding that exhibits better generalisation. This was demonstrated by noting that the permutation invariant IQP encoding accuracy scaled up as the number of points increased, whereas a non-invariant IQP encoding performed worse as the number of points increased. These findings suggest that this method may be a promising solution to the problem of QSVM generalisation performance worsening with increasing qubits \cite{inductive_bias, huang_power_2021}, and may have potential applications in future QML algorithms in various fields such as object recognition and particle physics. 

This work demonstrates an encoding method to improve generalisation for permutation invariant data, however, it does not necessarily guarantee a quantum advantage exists. Recently there have been results showing that qubit permutation invariant operators may be classically tractable under certain conditions due to their reduction in dimensionality \cite{anschuetz_efficient_2023}. A key difference between our work and methods that utilise equivariant variational models \cite{meyer22, Nguyen22, schatzki_theoretical_2022, Kazi23} is that the method presented in this paper implements permutation invariance directly into the encoding step, without considering the classification model. Hence, the variational model does not need to be permutation invariant in order to capture the symmetry. Further investigation into methods of efficient implementation of our technique could help improve the technique by overcoming problems such as the $\mathcal{O}(n^2)$ scaling of ancilla qubits in the probabilistic symmetrisation circuit \cite{barenco1996}.

Future work could focus on extending this encoding to other types of symmetries, such as rotational symmetry or translation symmetry, which could be relevant for point clouds and other data types such as images or time series. There is also the possible challenge of finding an efficient implementation on real devices and assessing how the encodings will perform in the presence of noise. The technique suggested here could be used for any data that exhibits permutation invariance, including cases where the input data itself is quantum. Real applications will be dependent on the rate of technological advancement of quantum machines, however, near-term use cases could focus on point clouds with a small number of points, as is often the case in particle physics data.

\backmatter

\bmhead{Acknowledgments}

This work was supported by the University of Melbourne through the establishment of an IBM Quantum Network Hub at the University. This research was supported by the Australian Research Council from grant DP210102831. JH acknowledges the support of the Research Training Program Scholarship and the N.D. Goldsworthy Scholarship. CDH is supported by a research grant from the Laby Foundation.

\section*{Declarations}

\bmhead{Data Availability}
The datasets generated during and/or analysed during the current study are available from the corresponding author on reasonable request.

\bmhead{Competing interests}
 On behalf of all authors, the corresponding author states that there is no conflict of interest.

\bibliography{sn-bibliography}% common bib file


\begin{thebibliography}{35}
\ifx \bisbn   \undefined \def \bisbn  #1{ISBN #1}\fi
\ifx \binits  \undefined \def \binits#1{#1}\fi
\ifx \bauthor  \undefined \def \bauthor#1{#1}\fi
\ifx \batitle  \undefined \def \batitle#1{#1}\fi
\ifx \bjtitle  \undefined \def \bjtitle#1{#1}\fi
\ifx \bvolume  \undefined \def \bvolume#1{\textbf{#1}}\fi
\ifx \byear  \undefined \def \byear#1{#1}\fi
\ifx \bissue  \undefined \def \bissue#1{#1}\fi
\ifx \bfpage  \undefined \def \bfpage#1{#1}\fi
\ifx \blpage  \undefined \def \blpage #1{#1}\fi
\ifx \burl  \undefined \def \burl#1{\textsf{#1}}\fi
\ifx \doiurl  \undefined \def \doiurl#1{\url{https://doi.org/#1}}\fi
\ifx \betal  \undefined \def \betal{\textit{et al.}}\fi
\ifx \binstitute  \undefined \def \binstitute#1{#1}\fi
\ifx \binstitutionaled  \undefined \def \binstitutionaled#1{#1}\fi
\ifx \bctitle  \undefined \def \bctitle#1{#1}\fi
\ifx \beditor  \undefined \def \beditor#1{#1}\fi
\ifx \bpublisher  \undefined \def \bpublisher#1{#1}\fi
\ifx \bbtitle  \undefined \def \bbtitle#1{#1}\fi
\ifx \bedition  \undefined \def \bedition#1{#1}\fi
\ifx \bseriesno  \undefined \def \bseriesno#1{#1}\fi
\ifx \blocation  \undefined \def \blocation#1{#1}\fi
\ifx \bsertitle  \undefined \def \bsertitle#1{#1}\fi
\ifx \bsnm \undefined \def \bsnm#1{#1}\fi
\ifx \bsuffix \undefined \def \bsuffix#1{#1}\fi
\ifx \bparticle \undefined \def \bparticle#1{#1}\fi
\ifx \barticle \undefined \def \barticle#1{#1}\fi
\bibcommenthead
\ifx \bconfdate \undefined \def \bconfdate #1{#1}\fi
\ifx \botherref \undefined \def \botherref #1{#1}\fi
\ifx \url \undefined \def \url#1{\textsf{#1}}\fi
\ifx \bchapter \undefined \def \bchapter#1{#1}\fi
\ifx \bbook \undefined \def \bbook#1{#1}\fi
\ifx \bcomment \undefined \def \bcomment#1{#1}\fi
\ifx \oauthor \undefined \def \oauthor#1{#1}\fi
\ifx \citeauthoryear \undefined \def \citeauthoryear#1{#1}\fi
\ifx \endbibitem  \undefined \def \endbibitem {}\fi
\ifx \bconflocation  \undefined \def \bconflocation#1{#1}\fi
\ifx \arxivurl  \undefined \def \arxivurl#1{\textsf{#1}}\fi
\csname PreBibitemsHook\endcsname

%%% 1
\bibitem{huang_power_2021}
\begin{barticle}
\bauthor{\bsnm{Huang}, \binits{H.-Y.}},
\bauthor{\bsnm{Broughton}, \binits{M.}},
\bauthor{\bsnm{Mohseni}, \binits{M.}},
\bauthor{\bsnm{Babbush}, \binits{R.}},
\bauthor{\bsnm{Boixo}, \binits{S.}},
\bauthor{\bsnm{Neven}, \binits{H.}},
\bauthor{\bsnm{McClean}, \binits{J.R.}}:
\batitle{Power of data in quantum machine learning}.
\bjtitle{Nature Communications}
\bvolume{12}(\bissue{1}),
\bfpage{2631}
(\byear{2021}).
\doiurl{10.1038/s41467-021-22539-9}
\end{barticle}
\endbibitem

%%% 2
\bibitem{biamonte_quantum_2017}
\begin{barticle}
\bauthor{\bsnm{Biamonte}, \binits{J.}},
\bauthor{\bsnm{Wittek}, \binits{P.}},
\bauthor{\bsnm{Pancotti}, \binits{N.}},
\bauthor{\bsnm{Rebentrost}, \binits{P.}},
\bauthor{\bsnm{Wiebe}, \binits{N.}},
\bauthor{\bsnm{Lloyd}, \binits{S.}}:
\batitle{Quantum machine learning}.
\bjtitle{Nature}
\bvolume{549}(\bissue{7671}),
\bfpage{195}--\blpage{202}
(\byear{2017}).
\doiurl{10.1038/nature23474}
\end{barticle}
\endbibitem

%%% 3
\bibitem{zeguendry_quantum_2023}
\begin{barticle}
\bauthor{\bsnm{Zeguendry}, \binits{A.}},
\bauthor{\bsnm{Jarir}, \binits{Z.}},
\bauthor{\bsnm{Quafafou}, \binits{M.}}:
\batitle{Quantum {Machine} {Learning}: {A} {Review} and {Case} {Studies}}.
\bjtitle{Entropy}
\bvolume{25}(\bissue{2}),
\bfpage{287}
(\byear{2023}).
\doiurl{10.3390/e25020287}
\end{barticle}
\endbibitem

%%% 4
\bibitem{sajjan_quantum_2022}
\begin{barticle}
\bauthor{\bsnm{Sajjan}, \binits{M.}},
\bauthor{\bsnm{Li}, \binits{J.}},
\bauthor{\bsnm{Selvarajan}, \binits{R.}},
\bauthor{\bsnm{Sureshbabu}, \binits{S.H.}},
\bauthor{\bsnm{Kale}, \binits{S.S.}},
\bauthor{\bsnm{Gupta}, \binits{R.}},
\bauthor{\bsnm{Singh}, \binits{V.}},
\bauthor{\bsnm{Kais}, \binits{S.}}:
\batitle{Quantum machine learning for chemistry and physics}.
\bjtitle{Chemical Society Reviews}
\bvolume{51}(\bissue{15}),
\bfpage{6475}--\blpage{6573}
(\byear{2022}).
\doiurl{10.1039/D2CS00203E}
\end{barticle}
\endbibitem

%%% 5
\bibitem{Heredge21}
\begin{botherref}
\oauthor{\bsnm{Heredge}, \binits{J.}},
\oauthor{\bsnm{Hill}, \binits{C.}},
\oauthor{\bsnm{Hollenberg}, \binits{L.}},
\oauthor{\bsnm{Sevior}, \binits{M.}}:
Quantum support vector machines for continuum suppression in b meson decays.
Computing and Software for Big Science
\textbf{5}
(2021).
\doiurl{10.1007/s41781-021-00075-x}
\end{botherref}
\endbibitem

%%% 6
\bibitem{tuysuz_hybrid_2021}
\begin{barticle}
\bauthor{\bsnm{Tüysüz}, \binits{C.}},
\bauthor{\bsnm{Rieger}, \binits{C.}},
\bauthor{\bsnm{Novotny}, \binits{K.}},
\bauthor{\bsnm{Demirköz}, \binits{B.}},
\bauthor{\bsnm{Dobos}, \binits{D.}},
\bauthor{\bsnm{Potamianos}, \binits{K.}},
\bauthor{\bsnm{Vallecorsa}, \binits{S.}},
\bauthor{\bsnm{Vlimant}, \binits{J.-R.}},
\bauthor{\bsnm{Forster}, \binits{R.}}:
\batitle{Hybrid quantum classical graph neural networks for particle track
  reconstruction}.
\bjtitle{Quantum Machine Intelligence}
\bvolume{3}(\bissue{2}),
\bfpage{29}
(\byear{2021}).
\doiurl{10.1007/s42484-021-00055-9}
\end{barticle}
\endbibitem

%%% 7
\bibitem{pregnolato_sars-cov-2_2023}
\begin{barticle}
\bauthor{\bsnm{Pregnolato}, \binits{M.}},
\bauthor{\bsnm{Zizzi}, \binits{P.}}:
\batitle{{SARS}-{CoV}-2 spike and {ACE2} entanglement-like binding}.
\bjtitle{Quantum Machine Intelligence}
\bvolume{5}(\bissue{1}),
\bfpage{8}
(\byear{2023}).
\doiurl{10.1007/s42484-023-00098-0}
\end{barticle}
\endbibitem

%%% 8
\bibitem{azevedo_quantum_2022}
\begin{barticle}
\bauthor{\bsnm{Azevedo}, \binits{V.}},
\bauthor{\bsnm{Silva}, \binits{C.}},
\bauthor{\bsnm{Dutra}, \binits{I.}}:
\batitle{Quantum transfer learning for breast cancer detection}.
\bjtitle{Quantum Machine Intelligence}
\bvolume{4}(\bissue{1}),
\bfpage{5}
(\byear{2022}).
\doiurl{10.1007/s42484-022-00062-4}
\end{barticle}
\endbibitem

%%% 9
\bibitem{Yuan:2022jcw}
\begin{botherref}
\oauthor{\bsnm{Yuan}, \binits{X.-J.}},
\oauthor{\bsnm{Chen}, \binits{Z.-Q.}},
\oauthor{\bsnm{Liu}, \binits{Y.-D.}},
\oauthor{\bsnm{Xie}, \binits{Z.}},
\oauthor{\bsnm{Jin}, \binits{X.-M.}},
\oauthor{\bsnm{Liu}, \binits{Y.-Z.}},
\oauthor{\bsnm{Wen}, \binits{X.}},
\oauthor{\bsnm{Tang}, \binits{H.}}:
{Quantum support vector machines for aerodynamic classification}
(2022)
{\href{https://arxiv.org/abs/2208.07138}{{arXiv:2208.07138}}}
\end{botherref}
\endbibitem

%%% 10
\bibitem{meichanetzidis_grammar-aware_2023}
\begin{barticle}
\bauthor{\bsnm{Meichanetzidis}, \binits{K.}},
\bauthor{\bsnm{Toumi}, \binits{A.}},
\bauthor{\bparticle{de} \bsnm{Felice}, \binits{G.}},
\bauthor{\bsnm{Coecke}, \binits{B.}}:
\batitle{Grammar-aware sentence classification on quantum computers}.
\bjtitle{Quantum Machine Intelligence}
\bvolume{5}(\bissue{1}),
\bfpage{10}
(\byear{2023}).
\doiurl{10.1007/s42484-023-00097-1}
\end{barticle}
\endbibitem

%%% 11
\bibitem{havlicek_supervised_2019}
\begin{barticle}
\bauthor{\bsnm{Havlíček}, \binits{V.}},
\bauthor{\bsnm{Córcoles}, \binits{A.D.}},
\bauthor{\bsnm{Temme}, \binits{K.}},
\bauthor{\bsnm{Harrow}, \binits{A.W.}},
\bauthor{\bsnm{Kandala}, \binits{A.}},
\bauthor{\bsnm{Chow}, \binits{J.M.}},
\bauthor{\bsnm{Gambetta}, \binits{J.M.}}:
\batitle{Supervised learning with quantum-enhanced feature spaces}.
\bjtitle{Nature}
\bvolume{567}(\bissue{7747}),
\bfpage{209}--\blpage{212}
(\byear{2019}).
\doiurl{10.1038/s41586-019-0980-2}
\end{barticle}
\endbibitem

%%% 12
\bibitem{liu_rigorous_2021}
\begin{barticle}
\bauthor{\bsnm{Liu}, \binits{Y.}},
\bauthor{\bsnm{Arunachalam}, \binits{S.}},
\bauthor{\bsnm{Temme}, \binits{K.}}:
\batitle{A rigorous and robust quantum speed-up in supervised machine
  learning}.
\bjtitle{Nature Physics}
\bvolume{17}(\bissue{9}),
\bfpage{1013}--\blpage{1017}
(\byear{2021}).
\doiurl{10.1038/s41567-021-01287-z}
\end{barticle}
\endbibitem

%%% 13
\bibitem{lisnichenko_quantum_2022}
\begin{barticle}
\bauthor{\bsnm{Lisnichenko}, \binits{M.}},
\bauthor{\bsnm{Protasov}, \binits{S.}}:
\batitle{Quantum image representation: a review}.
\bjtitle{Quantum Machine Intelligence}
\bvolume{5}(\bissue{1}),
\bfpage{2}
(\byear{2022}).
\doiurl{10.1007/s42484-022-00089-7}
\end{barticle}
\endbibitem

%%% 14
\bibitem{Anand2022}
\begin{botherref}
\oauthor{\bsnm{Anand}, \binits{A.}},
\oauthor{\bsnm{Lyu}, \binits{M.}},
\oauthor{\bsnm{Baweja}, \binits{P.S.}},
\oauthor{\bsnm{Patil}, \binits{V.}}:
Quantum Image Processing.
arXiv
(2022).
\doiurl{10.48550/arxiv.2203.01831}
\end{botherref}
\endbibitem

%%% 15
\bibitem{west2022}
\begin{botherref}
\oauthor{\bsnm{West}, \binits{M.}},
\oauthor{\bsnm{Sevior}, \binits{M.}},
\oauthor{\bsnm{Usman}, \binits{M.}}:
Reflection Equivariant Quantum Neural Networks for Enhanced Image
  Classification
(2022).
\doiurl{10.48550/arxiv.2212.00264}
\end{botherref}
\endbibitem

%%% 16
\bibitem{schuld_effect_2020}
\begin{barticle}
\bauthor{\bsnm{Schuld}, \binits{M.}},
\bauthor{\bsnm{Sweke}, \binits{R.}},
\bauthor{\bsnm{Meyer}, \binits{J.J.}}:
\batitle{Effect of data encoding on the expressive power of variational
  quantum-machine-learning models}.
\bjtitle{Phys. Rev. A}
\bvolume{103},
\bfpage{032430}
(\byear{2021}).
\doiurl{10.1103/PhysRevA.103.032430}
\end{barticle}
\endbibitem

%%% 17
\bibitem{inductive_bias}
\begin{botherref}
\oauthor{\bsnm{Kübler}, \binits{J.M.}},
\oauthor{\bsnm{Buchholz}, \binits{S.}},
\oauthor{\bsnm{Schölkopf}, \binits{B.}}:
The inductive bias of quantum kernels
(2021).
\doiurl{10.48550/arxiv.2106.03747}
\end{botherref}
\endbibitem

%%% 18
\bibitem{PhysRevA.106.042407}
\begin{barticle}
\bauthor{\bsnm{Shaydulin}, \binits{R.}},
\bauthor{\bsnm{Wild}, \binits{S.M.}}:
\batitle{Importance of kernel bandwidth in quantum machine learning}.
\bjtitle{Phys. Rev. A}
\bvolume{106},
\bfpage{042407}
(\byear{2022}).
\doiurl{10.1103/PhysRevA.106.042407}
\end{barticle}
\endbibitem

%%% 19
\bibitem{bowles_contextuality_2023}
\begin{botherref}
\oauthor{\bsnm{Bowles}, \binits{J.}},
\oauthor{\bsnm{Wright}, \binits{V.J.}},
\oauthor{\bsnm{Farkas}, \binits{M.}},
\oauthor{\bsnm{Killoran}, \binits{N.}},
\oauthor{\bsnm{Schuld}, \binits{M.}}:
Contextuality and inductive bias in quantum machine learning.
arXiv.
arXiv:2302.01365 [quant-ph]
(2023).
\url{http://arxiv.org/abs/2302.01365}
Accessed 2023-03-24
\end{botherref}
\endbibitem

%%% 20
\bibitem{Chen2021}
\begin{barticle}
\bauthor{\bsnm{Chen}, \binits{S.}},
\bauthor{\bsnm{Liu}, \binits{B.}},
\bauthor{\bsnm{Feng}, \binits{C.}},
\bauthor{\bsnm{Vallespi-Gonzalez}, \binits{C.}},
\bauthor{\bsnm{Wellington}, \binits{C.}}:
\batitle{3d point cloud processing and learning for autonomous driving:
  Impacting map creation, localization, and perception}.
\bjtitle{IEEE Signal Processing Magazine}
\bvolume{38}(\bissue{1}),
\bfpage{68}--\blpage{86}
(\byear{2021}).
\doiurl{10.1109/MSP.2020.2984780}
\end{barticle}
\endbibitem

%%% 21
\bibitem{Mikuni_2021}
\begin{barticle}
\bauthor{\bsnm{Mikuni}, \binits{V.}},
\bauthor{\bsnm{Canelli}, \binits{F.}}:
\batitle{Point cloud transformers applied to collider physics}.
\bjtitle{Machine Learning: Science and Technology}
\bvolume{2}(\bissue{3}),
\bfpage{035027}
(\byear{2021}).
\doiurl{10.1088/2632-2153/ac07f6}
\end{barticle}
\endbibitem

%%% 22
\bibitem{openAI_pointE}
\begin{botherref}
\oauthor{\bsnm{Nichol}, \binits{A.}},
\oauthor{\bsnm{Jun}, \binits{H.}},
\oauthor{\bsnm{Dhariwal}, \binits{P.}},
\oauthor{\bsnm{Mishkin}, \binits{P.}},
\oauthor{\bsnm{Chen}, \binits{M.}}:
Point-E: A System for Generating 3D Point Clouds from Complex Prompts.
arXiv
(2022).
\doiurl{10.48550/arxiv.2212.08751}
\end{botherref}
\endbibitem

%%% 23
\bibitem{pointnet}
\begin{botherref}
\oauthor{\bsnm{Qi}, \binits{C.R.}},
\oauthor{\bsnm{Su}, \binits{H.}},
\oauthor{\bsnm{Mo}, \binits{K.}},
\oauthor{\bsnm{Guibas}, \binits{L.J.}}:
PointNet: Deep Learning on Point Sets for 3D Classification and Segmentation.
arXiv
(2016).
\doiurl{10.48550/arxiv.1612.00593}
\end{botherref}
\endbibitem

%%% 24
\bibitem{Shi2020TrainingAQ}
\begin{bchapter}
\bauthor{\bsnm{Shi}, \binits{R.}},
\bauthor{\bsnm{Tang}, \binits{H.}},
\bauthor{\bsnm{Jin}, \binits{X.-m.}}:
\bctitle{Training a quantum pointnet with nesterov accelerated gradient
  estimation by projection}.
(\byear{2020}).
\bcomment{Presented at the First Workshop on Quantum Tensor Networks in Machine
  Learning, 34th Conference on Neural Information Processing Systems (NeurIPS
  2020).}
\burl{\url{https://tensorworkshop.github.io/NeurIPS2020/accepted_papers/simptnet_cameraready_v2.pdf}}
\end{bchapter}
\endbibitem

%%% 25
\bibitem{meyer22}
\begin{botherref}
\oauthor{\bsnm{Meyer}, \binits{J.J.}},
\oauthor{\bsnm{Mularski}, \binits{M.}},
\oauthor{\bsnm{Gil-Fuster}, \binits{E.}},
\oauthor{\bsnm{Mele}, \binits{A.A.}},
\oauthor{\bsnm{Arzani}, \binits{F.}},
\oauthor{\bsnm{Wilms}, \binits{A.}},
\oauthor{\bsnm{Eisert}, \binits{J.}}:
Exploiting symmetry in variational quantum machine learning.
arXiv
(2022).
\doiurl{10.48550/arxiv.2205.06217}
\end{botherref}
\endbibitem

%%% 26
\bibitem{Nguyen22}
\begin{botherref}
\oauthor{\bsnm{Nguyen}, \binits{Q.T.}},
\oauthor{\bsnm{Schatzki}, \binits{L.}},
\oauthor{\bsnm{Braccia}, \binits{P.}},
\oauthor{\bsnm{Ragone}, \binits{M.}},
\oauthor{\bsnm{Coles}, \binits{P.J.}},
\oauthor{\bsnm{Sauvage}, \binits{F.}},
\oauthor{\bsnm{Larocca}, \binits{M.}},
\oauthor{\bsnm{Cerezo}, \binits{M.}}:
Theory for Equivariant Quantum Neural Networks.
arXiv
(2022).
\doiurl{10.48550/arxiv.2210.08566}
\end{botherref}
\endbibitem

%%% 27
\bibitem{schatzki_theoretical_2022}
\begin{botherref}
\oauthor{\bsnm{Schatzki}, \binits{L.}},
\oauthor{\bsnm{Larocca}, \binits{M.}},
\oauthor{\bsnm{Nguyen}, \binits{Q.T.}},
\oauthor{\bsnm{Sauvage}, \binits{F.}},
\oauthor{\bsnm{Cerezo}, \binits{M.}}:
Theoretical Guarantees for Permutation-Equivariant Quantum Neural Networks
(2022).
\url{https://arxiv.org/abs/2210.09974}
\end{botherref}
\endbibitem

%%% 28
\bibitem{Kazi23}
\begin{botherref}
\oauthor{\bsnm{Kazi}, \binits{S.}},
\oauthor{\bsnm{Larocca}, \binits{M.}},
\oauthor{\bsnm{Cerezo}, \binits{M.}}:
On the universality of $S_n$-equivariant $k$-body gates
(2023).
\url{https://arxiv.org/abs/2303.00728}
\end{botherref}
\endbibitem

%%% 29
\bibitem{anschuetz_efficient_2023}
\begin{botherref}
\oauthor{\bsnm{Anschuetz}, \binits{E.R.}},
\oauthor{\bsnm{Bauer}, \binits{A.}},
\oauthor{\bsnm{Kiani}, \binits{B.T.}},
\oauthor{\bsnm{Lloyd}, \binits{S.}}:
Efficient classical algorithms for simulating symmetric quantum systems.
arXiv:2211.16998 [quant-ph]
(2023).
\doiurl{10.48550/arXiv.2211.16998}
\end{botherref}
\endbibitem

%%% 30
\bibitem{barenco1996}
\begin{botherref}
\oauthor{\bsnm{Barenco}, \binits{A.}},
\oauthor{\bsnm{Berthiaume}, \binits{A.}},
\oauthor{\bsnm{Deutsch}, \binits{D.}},
\oauthor{\bsnm{Ekert}, \binits{A.}},
\oauthor{\bsnm{Jozsa}, \binits{R.}},
\oauthor{\bsnm{Macchiavello}, \binits{C.}}:
Stabilisation of Quantum Computations by Symmetrisation.
arXiv
(1996).
\doiurl{10.48550/arxiv.quant-ph/9604028}
\end{botherref}
\endbibitem

%%% 31
\bibitem{qiskit}
\begin{botherref}
\oauthor{\bsnm{Abraham}, \binits{H.}},
\oauthor{\bsnm{AduOffei}},
\oauthor{\bsnm{Agarwal}, \binits{R.}},
\oauthor{\bsnm{Akhalwaya}, \binits{I.Y.}},
\oauthor{\bsnm{Aleksandrowicz}, \binits{G.}},
\oauthor{\bparticle{et~al} \bsnm{{\v{C}}epulkovskis}}:
Qiskit: An Open-source Framework for Quantum Computing
(2019).
\doiurl{10.5281/zenodo.2562110}
\end{botherref}
\endbibitem

%%% 32
\bibitem{buzek_optimal_2000}
\begin{barticle}
\bauthor{\bsnm{Buzek}, \binits{V.}},
\bauthor{\bsnm{Hillery}, \binits{M.}}:
\batitle{Optimal manipulations with qubits: {Universal} quantum entanglers}.
\bjtitle{Physical Review A}
\bvolume{62}(\bissue{2}),
\bfpage{022303}
(\byear{2000}).
\doiurl{10.1103/PhysRevA.62.022303}
\end{barticle}
\endbibitem

%%% 33
\bibitem{creevey23}
\begin{botherref}
\oauthor{\bsnm{Creevey}, \binits{F.M.}},
\oauthor{\bsnm{Hill}, \binits{C.D.}},
\oauthor{\bsnm{Hollenberg}, \binits{L.C.L.}}:
GASP -- A Genetic Algorithm for State Preparation
(2023).
\doiurl{10.48550/arxiv.2302.11141}
\end{botherref}
\endbibitem

%%% 34
\bibitem{zoufal_quantum_2019}
\begin{barticle}
\bauthor{\bsnm{Zoufal}, \binits{C.}},
\bauthor{\bsnm{Lucchi}, \binits{A.}},
\bauthor{\bsnm{Woerner}, \binits{S.}}:
\batitle{Quantum {Generative} {Adversarial} {Networks} for learning and loading
  random distributions}.
\bjtitle{npj Quantum Information}
\bvolume{5}(\bissue{1}),
\bfpage{1}--\blpage{9}
(\byear{2019}).
\doiurl{10.1038/s41534-019-0223-2}
\end{barticle}
\endbibitem

%%% 35
\bibitem{park21}
\begin{botherref}
\oauthor{\bsnm{Park}, \binits{C.-Y.}}:
Efficient ground state preparation in variational quantum eigensolver with
  symmetry breaking layers.
arXiv
(2021).
\doiurl{10.48550/arxiv.2106.02509}
\end{botherref}
\endbibitem

\end{thebibliography}


\clearpage

\begin{appendices}

\section{Implementation}

The methodology employed in this study utilises the quantum support vector machine (QSVM) approach. The QSVM technique is based on the encoding of classical data, denoted by $X$, into a higher-dimensional space to perform classification tasks. The encoding circuit $\mathcal{U}$ establishes a mapping $\mathcal{U}: X \xrightarrow{} \ket{\psi(X)}\bra{\psi(X)}$, which transforms classical data $X$ into a quantum state represented by a density matrix $\ket{\psi(X)}\bra{\psi(X)}$. Importantly, it is not necessary to explicitly define the higher-dimensional encoding $\ket{\psi(X)}\bra{\psi(X)}$, only the inner product between data points in the higher-dimensional space, $ \rvert \langle \psi(X_i) \rvert \psi(X_j) \rangle \rvert ^2$. The inner product between two points $i$ and $j$ becomes entry $K_{i, j}$ of the kernel matrix $K$. Consequently, the actual device implementation consists of calculating the overlap between two quantum states to approximate the kernel matrix, which is determined by measuring the quantity $K_{i,j} = \rvert \langle 0 \rvert \mathcal{U}(X_i)\mathcal{U}^\dagger(X_j) \rvert 0 \rangle \rvert ^2$. Therefore, an encoding circuit $\mathcal{U}$ is required, involving the proposed permutation symmetrisation step and its conjugate version.

For the simulation implementation, we employed the Qiskit $\textit{statevector\_simulator}$ to compute the kernel using mathematical methods for determining the exact symmetric states. A point cloud $X$ contains $n$ points, with the coordinates of point $i$ given by $\textbf{p}_i$. In our encoding circuit $\mathcal{U}$, each point $\textbf{p}_i$ is initially encoded into a quantum state using the point encoding circuit $U: \textbf{p}_i \xrightarrow{} \ket{\textbf{p}_i}$. As the $\textit{statevector\_simulator}$ was utilised in this process, the amplitudes of all $\ket{\textbf{p}_i}$ are known. Thus, we can compute a permutation by finding the tensor product of all point quantum states. We can subsequently sum over all $n!$ possible permutations to derive the permutation invariant statevector for the entire point cloud $\ket{X_s} = \mathcal{N}\sum{\sigma \in S_n}\ket{\textbf{p}_{\sigma_1}}\ket{\textbf{p}_{\sigma_2}}...\ket{\textbf{p}_{\sigma_n}}$ and normalising the state. Consequently, it is feasible to directly compute the quantity $\rvert \langle \psi(X_i) \rvert \psi(X_j) \rangle \rvert ^2$ for any two point clouds. This, in turn, allows for the calculation of the entire kernel matrix for a dataset in a manner that is simpler to calculate classically than simulating entire circuits. This algorithm underpins the theoretical results presented in this paper. The process of generating the permutation invariant state is summarised in Algorithm~\ref{algo1}.

In this study, we employed the QSVM method to concentrate solely on the encoding step, without considering any variational ansatz structure. However, in practice, using the real circuit implementation discussed in Section~\ref{sec_circuit}, the QSVM approach may prove to be rather inefficient due to the necessity of constructing two permutation invariant states simultaneously when computing the kernel entries if using this method. This means that in practice it may be more efficient to use a variational ansatz circuit, placed after the encoding circuit.

\section{Distribution Specifications}

The torus distribution was generated using the following
\begin{equation}
    x = \Big(1 + \cos(s) \Big) \cos(t)
\end{equation}
\begin{equation}
    y = \Big(1 + \cos(s) \Big) \sin(t)
\end{equation}
\begin{equation}
    z = \sin(s).
\end{equation}
The average magnitude of the points in this distribution was then calculated to be approximately $1.28$. We then scaled down the distribution by this factor so that the average magnitude was $1$, matching the radius of the sphere.

\section{Probabilistic Symmetric Encoding}

Here we shall discuss the scaling of the worst-case probability of being able to prepare a symmetric states using the circuit suggested by Barenco et al. \cite{barenco1996}. Looking back at the two-qubit example shown in Figure~\ref{fig_2_qubit_example} we start with a control qubit in the state $\frac{1}{\sqrt{2}}(\ket{0} + \ket{1})$. This is the control qubit for a controlled swap gate that acts on the two initial states $\ket{\textbf{p}_1}$ and $\ket{\textbf{p}_2}$ resulting in the following state

\begin{equation}\label{2_qubit_intermediate_state}
    \frac{1}{\sqrt{2}}(\ket{0}\ket{\textbf{p}_1}\ket{\textbf{p}_2} + \ket{1}\ket{\textbf{p}_2}\ket{\textbf{p}_1}).
\end{equation}

The procedure then applies a Hadamard gate to the control qubit prior to measurement in the Z basis. This takes the state to the following
\begin{align}
\begin{split}\label{equation_2_qubit_sym}
    \ket{\psi} = \frac{1}{2}\ket{0}(\ket{\textbf{p}_1}\ket{\textbf{p}_2} + \ket{\textbf{p}_2}\ket{\textbf{p}_1}) + \\
    \frac{1}{2}\ket{1}(\ket{\textbf{p}_1}\ket{\textbf{p}_2} - \ket{\textbf{p}_2}\ket{\textbf{p}_1})
\end{split}
\end{align}

From here the probability of measuring the ancilla qubit in the $\ket{0}$ state is given by applying the probability operator $P(0) = \bra{\psi} \Big( \ket{0}\bra{0} \otimes I \otimes I \Big) \ket{\psi}$. This can be written as
\begin{equation}
     \frac{1}{4} \Big(\bra{\textbf{p}_1}\bra{\textbf{p}_2} + \bra{\textbf{p}_2}\bra{\textbf{p}_1}\Big) \Big(\ket{\textbf{p}_1}\ket{\textbf{p}_2} + \ket{\textbf{p}_2}\ket{\textbf{p}_1}\Big), 
\end{equation}
which can then be evaluated to
\begin{equation}
     \frac{1}{2}\Big( 1 + \rvert \langle \textbf{p}_1 \rvert \textbf{p}_2 \rangle \rvert^2 \Big).   
\end{equation}
Note that the lowest probability this can reach is $\frac{1}{2}$ in the case where $\rvert \langle \textbf{p}_1 \rvert \textbf{p}_2 \rangle \rvert = 0$. Once we measure $\ket{0}$ in the ancilla qubit we will collapse the state into the symmetric superposition, which is invariant under qubit permutation. We can write this state as 
\begin{equation*}
    \ket{X_s} = \mathcal{N}(\ket{\textbf{p}_1}\ket{\textbf{p}_2} + \ket{\textbf{p}_2}\ket{\textbf{p}_1}).
\end{equation*}

In order to calculate the normalisation constant $\mathcal{N}$, just note that $\rvert \langle X_s \rvert X_s \rangle \rvert = 1$ and from this we can calculate

\begin{equation}
    \rvert \mathcal{N} \rvert^2 (2 +  2 \rvert \langle \textbf{p}_1 \rvert \textbf{p}_2 \rangle \rvert^2) = 1,
\end{equation}

\begin{equation}
    \mathcal{N} = \frac{1}{\sqrt{2(1 +  \rvert \langle \textbf{p}_1 \rvert \textbf{p}_2 \rangle \rvert^2)}}.
\end{equation}

\section{Probabilistic Symmetric Ancilla Preparation}

In the method proposed by Barenco et al. \cite{barenco1996} one of the steps involves applying a unitary $V_f$ that prepares $f$ ancilla qubits into the state
\begin{equation}
    \frac{1}{\sqrt{f + 1}}(\ket{00...0} + \ket{10...0} + \ket{01...0} + ... + \ket{00...1})
\end{equation}
In order to create this state, we need to define two unitary operators
\begin{equation}
    R_f = \frac{1}{\sqrt{f+1}}
    \begin{pmatrix}
    1 & -\sqrt{f}\\
    \sqrt{f} & 1 
    \end{pmatrix},
\end{equation}

\begin{align}
\begin{split}
    T_f,j = 
    \begin{pmatrix}
    1 & 0 & 0 & 0\\
    0 & \frac{1}{\sqrt{f-j+1}} & \frac{\sqrt{f - j}}{\sqrt{f-j+1}} & 0 \\
    0 & -\frac{\sqrt{f - j}}{\sqrt{f-j+1}} & \frac{1}{\sqrt{f-j+1}} & 0\\
    0 & 0 & 0 & 1
    \end{pmatrix}.
\end{split}
\end{align}

If we have $f$ control qubits initially in the $\ket{0}^{\otimes f}$ state. Then we can prepare the desired state by applying $R_f$ to the first qubit and $T_f,j$ to adjacent qubits starting from $j=1$ to $j=f-1$. This entire procedure creates the unitary operation $V_f$. This is repeated for $f = 1,2,..,n-1$ where we have $n$ qubits in total to symmetrise, resulting in a total of $c=\frac{1}{2}n(n-1)$ ancilla qubits used.

\section{Success Probability for Symmetric Preparation}\label{apd_success_prob}

We present a small study of the probability of success when using the probabilistic technique for preparing permutation symmetric states described by Barenco et al. \cite{barenco1996}. We use angle encoding with random inputs to generate initial states with the results shown in Table~\ref{sym_circuit_table}.

\begin{table}[h]
 \caption{The average probability of producing permutation symmetrised states when given input states that were randomly initialised using angle encoding via an X rotation gate \cite{barenco1996}. This corresponds to the probability of measuring the ancilla qubits in the state $\ket{0}^{\otimes c}$. Each circuit was run over 10,000 shots using Qiskit $\textit{qasm\_simulator}$ and the proportion of $\ket{0}^{\otimes c}$ measured. This was then repeated over 1000 separate runs, using randomly initialised input states each time, reporting the average probability found over all runs. Note that the lower bound for the probabilities is much lower, but it is unlikely to be reached with input states randomly initialised via angle encoding.}\label{sym_circuit_table}%
\begin{center}
\begin{tabular}{|l|l|}
\hline
 Initial States $n$ & Mean Probability\\
 \hline
 2 &  0.842\\
 3 &  0.752\\
 4 &   0.630 \\
 5 &  0.530\\
\hline
\end{tabular}
\end{center}
\end{table}

The lowest values of probability are found when initialising input states that are maximally orthogonal to each other. Note that in practice the mean probability was found to be far higher than this when using random initial states, due to the fact that random states are unlikely to be maximally mutually orthogonal. Note that if basis state encoding was used, instead of angle encoding, then the possibility of orthogonal states would potentially be much higher, and thus the chance of successfully symmetrising the states significantly lower.

\end{appendices}

\end{document}